\documentclass[letterpaper, 10 pt, conference]{ieeeconf}  

\IEEEoverridecommandlockouts                              

\usepackage{graphics} 
\usepackage{epsfig} 
\usepackage{mathptmx} 
\usepackage{times} 
\usepackage{amsmath} 
\usepackage{amssymb}  
\usepackage{algorithm}
\usepackage{mathtools}

\usepackage{subfig} 
\usepackage{graphicx}
\usepackage[normalem]{ulem}

\usepackage{todonotes}

\newtheorem{remark}{\textbf{Remark}}

\usepackage{braket}

\title{\LARGE \bf
Learning Quadrotor Dynamics Using Neural Network for Flight Control
}

\author{Somil Bansal$^{*}$ \and Anayo K. Akametalu$^{*}$ \and Frank J. Jiang \and Forrest Laine  \and Claire J. Tomlin  
\thanks{The authors are with the Department of Electrical Engineering and Computer Sciences, 
        University of California, Berkeley, CA 94720.
        {\tt\small \{somil, kakametalu, forrest.laine, fjiang6o2, tomlin\}@eecs.berkeley.edu }}%
	\thanks{$^{*}$Both authors contributed equally to this work. This work is supported by the NSF CPS project ActionWebs under grant number
0931843, NSF CPS project FORCES under grant number 1239166, and by ONR under the HUNT, SMARTS and Embedded Humans MURIs, and by AFOSR under the CHASE MURI. The research of A.K. Akametalu has received funding from the UC Berkeley Chancellor's Fellowship.}%
}

\begin{document}

\maketitle
\thispagestyle{empty}
\pagestyle{empty}

\begin{abstract}
Traditional learning approaches proposed for controlling quadrotors or helicopters have focused on improving performance for specific trajectories by iteratively improving upon a nominal controller, for example learning from demonstrations, iterative learning, and reinforcement learning. In these schemes, however, it is not clear how the information gathered from the training trajectories can be used to synthesize controllers for more general trajectories. Recently, the efficacy of deep learning in inferring helicopter dynamics has been shown. Motivated by the generalization capability of deep learning, this paper investigates whether a neural network based dynamics model can be employed to synthesize control for trajectories different than those used for training. To test this, we learn a quadrotor dynamics model using only translational and only rotational training trajectories, each of which can be controlled independently, and then use it to simultaneously control the yaw and position of a quadrotor, which is non-trivial because of nonlinear couplings between the two motions. We validate our approach in experiments on a quadrotor testbed.   
\end{abstract}

\section{Introduction}
System identification, the mathematical modeling of a system's dynamics, is one of the most basic and important components of control. Constructing an appropriate model is often the first step in designing a controller. Modeling accuracy, therefore, directly impacts controller success and performance, as inaccuracies in the model appear to the controller as external disturbances. 

Quadrotors have recently emerged as a popular platform for unmanned aerial vehicle (UAV) research, due to the simplicity of their construction and maintenance. Quadrotors can be highly maneuverable, and have the potential to hover, take off, fly and land in small areas due to a vertical take off and landing (VTOL) capability \cite{hoffmann2007}. A quadrotor has four rotors located at the  four corners of a cross frame, and is controlled by changing the speed of rotation of the four rotors \cite{brookner1998tracking}, \cite{wan2000unscented}. However, the system is under-actuated, nonlinear and difficult to control on aggressive trajectories. Most of the work in this area focuses on designing controllers that are derived from a linearization of the model around hover conditions and are stable only under reasonably small roll and pitch angles. While advanced control methods such as feedback linearization \cite{voos2009nonlinear}, adaptive control \cite{nicol2011robust}, sliding-mode control \cite{xu2006sliding}, $H_{\infty}$ robust control \cite{raffo2010integral} have been developed, the performance of these control schemes depends heavily on the underlying model. In \cite{hoffmann2007}, the authors present an in-depth study of some of the advanced aerodynamics effects that can affect quadrotor flight, like blade flapping and effect of airflow. These effects, however, are hard to model and hence difficult to take into account while designing a controller. 
To circumvent these modeling issues, data-driven, learning-based control schemes have also been proposed (see \cite{zhaowei2015iterative}, \cite{hehn2013iterative} and references therein). An interesting approach has been presented in \cite{abbeel2} to successfully perform advanced aerobatics on a helicopter under autonomous control using apprenticeship learning. In this approach, a helicopter is flown on a trajectory repeatedly, and a target trajectory for control and time-varying dynamics are estimated from them. Together, these trajectories allow for successful control of the helicopter through advanced aerobatics. 
One limitation of the approaches above is that they are limited to designing a controller for specific trajectories; for a new trajectory, one has to learn the controller again from scratch.

\begin{figure}
  \centering
  \includegraphics[width=0.45\textwidth]{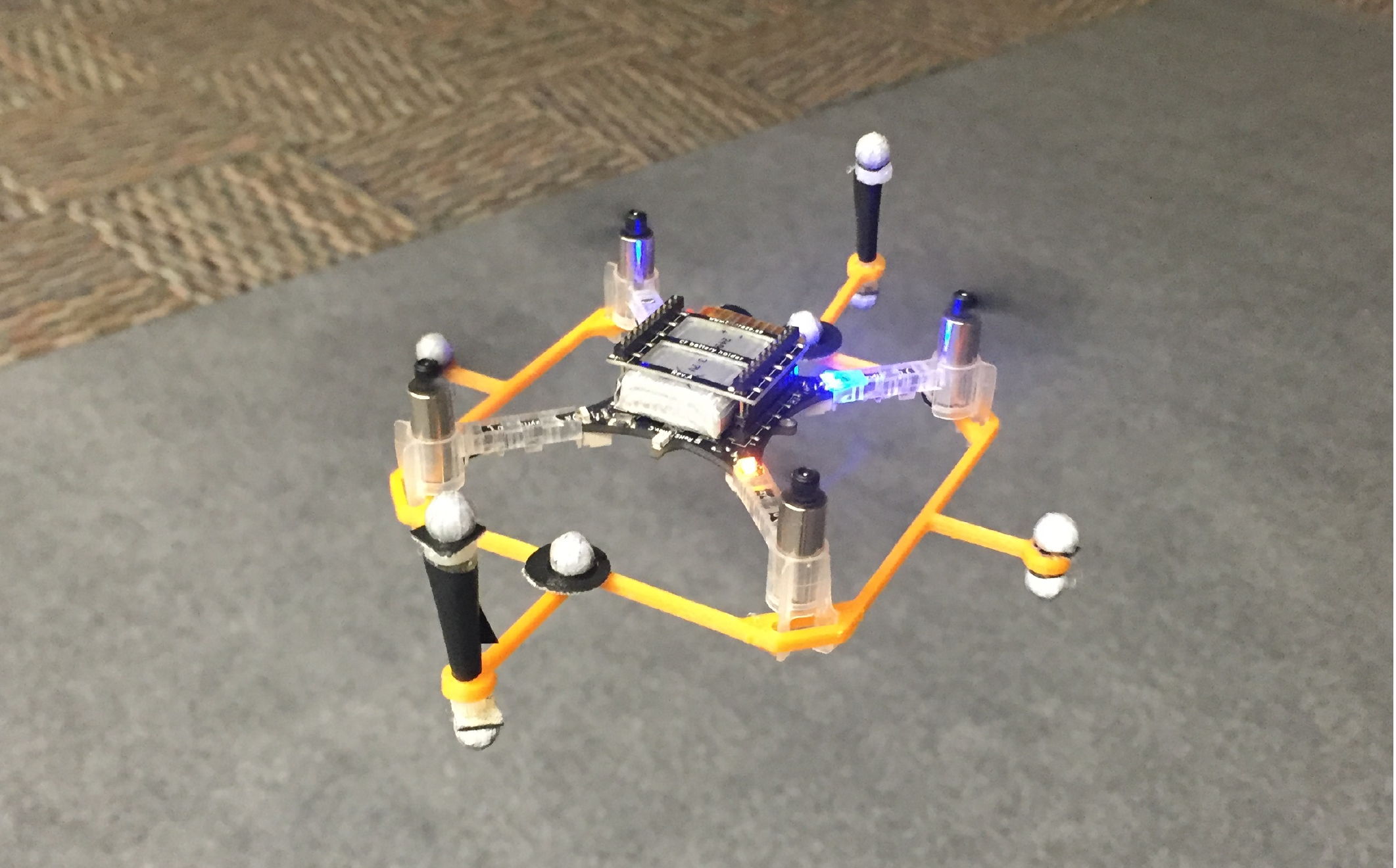}
  \caption{A picture of Crazyflie 2.0 quadrotor flying during one of our experiments.}
  \label{fig:crazyflie_pic}
  \vspace{-2em}
\end{figure}

The apprenticeship learning approach, however, indicates that the difficulty in modeling helicopter dynamics does not come from stochasticity in the system or unstructured noise in the demonstrations \cite{abbeel2}. Rather, the presence of unobserved states causes simple models to be inaccurate, even though repeatability in the system dynamics is preserved across repetitions of the same maneuver. One can thus use system data to model these dynamics directly in the entire state space rather than for specific trajectories. 

One potential approach can be to model such dynamics using neural networks. Neural networks (NN) are known to be universal function approximators; their structure allows them to model highly nonlinear functions and unobserved states directly from the observed data, which might in general be hard to model directly \cite{lecun2015deep}. Moreover, they can learn a generalized model that can be extended beyond the observed data. Motivated by this, the authors in \cite{punjanideep} propose a NN model to learn local unmodeled dynamics for a helicopter in different parts of the state space; thus, one need not learn the unmodeled dynamics for a specific trajectory. However, it is not clear if the proposed NN-based model can be used to control the system, and if the learned dynamics accurately represent the system beyond the data it was trained on.
%
In this paper, we answer these practically important questions and investigate (i) whether a highly nonlinear dynamics model given by a NN can be effectively used to design a controller for a quadrotor and (ii) whether it is general enough to be used to design a controller for the trajectories that the network was \textit{not} trained on. 

For this purpose, we collect state-input data of a nano-quadrotor Crazyflie 2.0 by flying it on the trajectories that consist of translational or rotational motion, but not both.
We next train a feed-forward Rectified-Linear Unit (ReLU) NN to learn the state-space dynamics of Crazyflie. To test the generalization capabilities of the trained NN, we use the learned NN model to control the quadrotor on a trajectory that consists of a simultaneous translational and rotational motion. A non-zero yaw angle introduces highly nonlinear couplings in the rotational and translational dynamics, and is the primary motivation behind why the quadrotor position control is studied generally regulating yaw to zero and vice-versa \cite{beard2008quadrotor, sanchez}. Thus, to successfully perform such a motion, the NN needs to infer these couplings from the individual translational and rotational trajectories it was trained on, and yet the model should be simple enough to design a controller. Our main contributions are:
\begin{itemize}
\item learning the dynamics of a quadrotor using a NN that is simple enough to be used for control purposes, but complex enough to accurately model system dynamics;
\item demonstrating that the current state-input data is sufficient to learn the dynamics to a good accuracy;
\item showing that NN can generalize the dynamics to learn nonlinear couplings between translational and rotational motions, even when the training data does not capture these couplings significantly; thus, the NN model can be used to fly the trajectories it was not trained on.
\end{itemize}  
%

\section{Quadrotor System Identification} \label{secn:sys_dyn}
In this section, we introduce our general quadrotor model and formulate the system identification problem for a quadrotor system. Consider a dynamical system with state vector $s$ and control inputs $u$. The goal of the system identification process is to find a function $f$ which maps from state-control space to state-derivative:
\begin{equation*}
\dot{s} = f(s, u; \alpha),
\end{equation*}
where the system model is parameterized by $\alpha$. The system identification task then becomes to find, given input and state data, parameters $\alpha$ that minimize the prediction error. Note that for a physics-based model, $\alpha$ generally captures the physical properties of the system (for example, mass, moment of inertia, etc. for a quadrotor); however, for a NN based model, the parameters can be thought of as degrees of freedom a NN has to learn different nonlinear function.  

The quadrotor system is modeled as a rigid body with a twelve dimensional state vector $s\coloneqq \begin{bmatrix} p & v &\zeta & \omega \end{bmatrix}$, which includes  the position $p=(x,y,z)$ in a North-East-Down inertial reference frame $I$, linear velocities $v=(\dot{x},\dot{y}, \dot{z})$ in $I$, attitude (orientation) represented by Euler angles $\zeta =(\phi,\theta, \psi)$, and angular velocities $\omega=(\omega_x,\omega_y,\omega_z)$ expressed in the body-fixed coordinate frame $B$ of the quadrotor. The Euler angles parameterize the coordinate transformation from $I$ to $B$ with the standard \emph{yaw-pitch-roll} convention, i.e. a rotation by $\psi$ about the $z$-axis in the inertial frame, followed by a rotation of $\theta$ about the $y$-axis of the body-fixed frame, and finally another rotation of $\phi$ about the $x$-axis in the new body-fixed frame. This is written compactly as
\begin{equation} \label{eq:fv}
_I^BR(\phi,\theta,\psi)=R_x(\phi)R_y(\theta)R_z(\psi),
\end{equation}
\noindent where $R_x$,$R_y$, and $R_z$ are basic $3\times 3$ rotation matrices about their respective axes.

The system is controlled via four inputs $u \coloneqq \begin{bmatrix}u_1 & u_2 & u_3 & u_4\end{bmatrix}$, where $u_1$ is the thrust along the $z$-axis in $B$, and $u_2$, $u_3$ and $u_4$ are rolling, pitching  
and yawing moments respectively, all in $B$ \footnote{These inputs are generated by varying the angular speeds of the four propellers, which map linearly to the inputs.}. The system evolves according to dynamics:
\begin{equation} \label{sysDynamics}
\dot{s} = \begin{bmatrix} \dot{p} \\ \dot{v} \\ \dot{\zeta} \\ \dot{\omega}\end{bmatrix} = f(s, u; \alpha) = \begin{bmatrix} v \\ f_v(s,u; \alpha_1) \\ \hat{R}\omega \\ f_{\omega}(s,u; \alpha_2) \end{bmatrix},
\end{equation}
where the system model is parameterized by $\alpha := (\alpha_1, \alpha_2)$. In Section \ref{ReLU}, we explain how $f_v$ and $f_{\omega}$ are exactly parameterized by $\alpha_1$ and $\alpha_2$, and how we can determine these parameters.
Note that $\dot{\zeta} \neq \omega$ in general. $\dot{\zeta}$, or Euler rates as they are called, can be obtained by rotating the angular velocities to the inertial frame \cite{beard2008quadrotor, hoffmann2007} and are given by:
\begin{equation} \label{eq:rpytobody}
\dot{\zeta} =\hat{R}\omega, \quad \hat{R}=\begin{bmatrix} 1 & \sin\phi \tan \theta  & \cos \phi \tan \theta \\ 0 & \cos \phi & -\sin \phi \\ 0 & \frac{\sin \phi}{\cos \theta} & \frac{\cos \phi}{\cos \theta} \end{bmatrix}.
\end{equation}
The unknown components in \eqref{sysDynamics} are $f_v$ and $f_{\omega}$, the linear (or translational) and angular (or rotational) acceleration that the quadrotor undergoes, which we aim to approximate with a NN as a function of state, control, and model parameters. The system identification task for the quadrotor is thus to determine $\alpha_1$ (resp. $\alpha_2$), given observed values of $f_v$ (resp. $f_{\omega}$), $s$, and $u$. In this work, we minimize mean squared prediction error (MSE) over a training set of collected data, solving
\begin{equation} \label{sysID_prob}
\min_{\alpha_1} \sum_{t=1}^T \frac{1}{T} \|\tilde{f}_{v, t} - f_v(s_t, u_t; \alpha_1) \|^2, 
\end{equation}
where $\tilde{f}_{v, t}$ are the observed values of $f_v$. A similar optimization problem can be defined for $f_{\omega}$. Depending on the forms of $f_v$ and $f_{\omega}$, \eqref{sysID_prob} results in a linear or a nonlinear least squares problem.
%


\section{Neural Network Model}\label{ReLU}
In this section, we present a neural network architecture to solve the system identification problem in \eqref{sysID_prob} and compute the parameters $\alpha_1, \alpha_2$ that minimize the MSE between predicted and observed data. 

As more and more data is being produced, and more and more computational power continues to become available, an important opportunity lies in harnessing data towards autonomy. In recent years, the fields of computer vision and speech processing have not only made significant leaps forward, but also rapidly increased their rate of progress, largely thanks to developments in deep learning \cite{krizhevsky2012imagenet, lecun2015deep}. Thus far the impact of deep learning has largely been in supervised learning. In supervised learning, each (training) example is a pair consisting of an input value (e.g., images) and a desired output value (e.g., `cat', `dog', etc. depending on what is in the image). After learning on the training data, the system is expected to make correct predictions for future (unseen) inputs. Supervised learning can thus also be thought as a direct high dimensional regression (or classification). 

\begin{figure}
  \centering
  \includegraphics[width=0.5\textwidth]{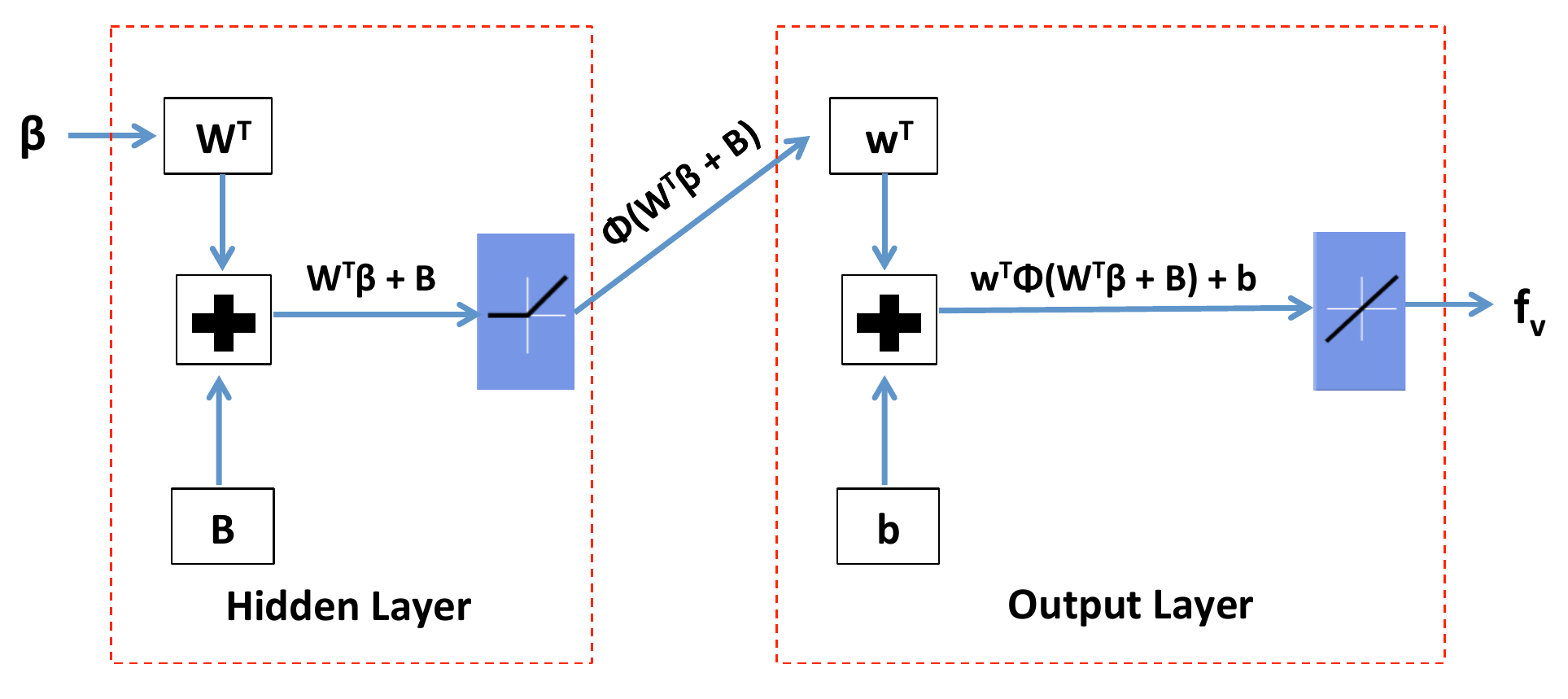}
  \caption{The neural network architecture used to learn $f_v$. The NN consists of two layers, a hidden ReLU layer and an output layer. The parameters to be learned during the training process are $\alpha_1 = (W, w, B, b)$. A similar architecture was used to learn $\alpha_2$ (and hence $f_{\omega}$).}
  \vspace{-1em}
  \label{fig:NNarch}
\end{figure}

Motivated by these advances, we train a multiple layer NN (i.e., ``deep learn" a NN) using supervised learning to predict the next state of the system based on current state and input. Our design is motivated by \cite{punjanideep}, wherein the authors deep learn the helicopter dynamics with a Rectified Linear Unit (ReLU) Network Model. A ReLU network model is a two-layer NN, consisting a hidden layer and an output layer, where the rectified-linear transfer function is used in the hidden layer. Algebraically, the model can be written as:

\begin{equation} \label{eqn:nnet}
f_v(\beta; \alpha_1) := w^{T} \phi(W^{T}\beta+ B) +b, 
\end{equation}

\noindent where $f_v$ represents the unknown linear acceleration component in \eqref{sysDynamics}, which is modeled by a NN
whose input is given by $\beta := (s,u) \in \mathbb{R}^{|\beta|}$. The NN has a hidden layer with $N$ units with weight matrix $W \in \mathbb{R}^{|\beta| \times N}$ and bias vector $B \in \mathbb{R}^{N}$, and a linear output layer of 3 units with weight matrix $w \in \mathbb{R}^{N \times 3}$ and bias vector $b \in \mathbb{R}^{3}$. $\phi$ represents the activation (or transfer) function of hidden units (also called ReLU activation function) and is given by $\phi(\cdot) = max(0, \cdot)$. The architecture of the NN is presented in Figure \ref{fig:NNarch}, which can be interpreted as follows: the input layer takes in the current state and input of the system. Each of $N$ hidden units computes the inner product of $\beta$ and one of the columns of $W$. The hidden units add a bias $B$ to the inner product and rectify this value at zero. The output layer is a linear combination of the hidden units, plus a final bias $b$. Intuitively, each hidden unit linearly partitions the input space into two parts based on $W$ and $B$. In one part the unit is inactive with zero output, while in the other it is active with positive output. Together, all hidden units partition the state space into polytopes. In each of these polytopes, the model has flexibility to learn the local dynamics.  

The goal of the training process is to determine (or ``learn") the parameters $\alpha_1 := (W, B, w, b)$ that minimize the MSE between the predicted acceleration $f_v$ and the observed acceleration $\tilde{f}_v$ subject to \eqref{eqn:nnet}. A similar NN architecture is used to learn $f_{\omega}$. Once the training process is complete, we can obtain a model for $f_v$ and $f_{\omega}$ by plugging in the optimal $\alpha_1$ and $\alpha_2$, obtained during training, in \eqref{eqn:nnet}. We, however, defer the exact details of the used hyperparameters and the training process until section \ref{secn:NN_training}.

\begin{remark}
Note that we only feed the current state and input in the network, and not any information on the past states and inputs unlike \cite{punjanideep}. Although giving the past state-input information will allow the NN to learn a more complex (and potentially more accurate) system dynamics model, it will also make it harder to design a controller for the resultant dynamics. So a simple input structure is chosen to make sure that the NN can be effectively employed to design a feedback controller. 
\end{remark}

\section{Control Design}  \label{secn:CD}
In this section, we aim to design a controller for the quadrotor system in \eqref{sysDynamics} to stabilize it on complex trajectories that involve both rotational and translational motions, such as a sinusoid-yaw trajectory (i.e., a trajectory where a quadrotor is flying on a sinusoid in the position coordinates (for example XY plane) while also yawing). 

In general, in a trajectory tracking problem, it may be impossible to exactly track a given desired trajectory due to limits imposed by the constraints on the system or when the trajectory is not dynamically feasible. This can happen, for example, for complex high dimensional systems such as quadrotors where it is relatively straightforward to specify the desired position and angular trajectories, but non-trivial to specify the linear and angular velocities such that the overall trajectory satisfies the system dynamics. It is therefore common practice to first compute a reference trajectory, which is the closest trajectory to the given desired trajectory satisfying system dynamics, and then, the optimal reference trajectory is tracked instead \cite{Rawlings2009}. We discuss the reference trajectory calculation in section \ref{secn:trajopt}. The stabilization of quadrotor on the reference trajectory is discussed in \ref{secn:LQR}. 


\subsection{Computation of a Feasible Reference}  \label{secn:trajopt}
Once we have computed $f_v$ and $f_{\omega}$ during the NN training phase, the full model of the quadrotor can be obtained from \eqref{sysDynamics}. Our goal in this section is to compute a dynamically feasible reference given a desired trajectory and the full system model \eqref{sysDynamics}. Since the system is controlled at discrete time points in our experiments, for ease of presentation we consider a discrete time approximation of \eqref{sysDynamics}:

\begin{equation} \label{eqn:sysDDynamics}
s(n+1) = s(n) + f(s(n),u(n);\alpha)\Delta t,
\end{equation}

\noindent where $n$ indexes the time step, $\Delta t$ is the sampling rate, $s(n)$ and $u(n)$ are the state and input of the quadrotor at time $n\Delta t$, and $\alpha := (\alpha_1, \alpha_2)$ are the parameters learned during the NN training. Given a horizon $N_H$ and a desired trajectory over that horizon ${\bf s}_d^{N_H} \coloneqq \{s_d(0),s_d(1),\hdots,s_{d}(N_H)\}$ our goal is to find a control signal  ${\bf u}^{N_H} \coloneqq \{ u(0),u(1),\hdots,u(N_H)\}$ that will achieve the desired trajectory when applied to the quadrotor.

In most cases the desired trajectory may not be dynamically feasible, so no such control signal exists. Instead, we look for a dynamically feasible trajectory that is ``as close as possible" to the desired trajectory. We thus want to solve the following optimization problem:

\begin{equation} \label{eq:trajopt}
\begin{aligned}
& \underset{{\bf s}^{N_H},{\bf u}^{N_H}}{\text{argmin}}
& & \sum_{n=0}^{N_H} \|s(n)-s_d(n)\|_2 \\
& \text{s. t.}
& & s(n+1)-s(n)=f(s(n),u(n); \alpha)\Delta t , \; n = 0, \ldots, N_H-1
\end{aligned}
\end{equation}

In words, we want to find the trajectory that minimizes the Euclidean distance to the desired trajectory, and the control that achieves such a trajectory. 
Since the NN output \eqref{eqn:nnet} is nonlinear, $f$ is nonlinear; therefore, the above optimization problem is a non-convex problem. In this paper, we use the sequential convex optimization (SCP) procedure proposed in \cite{schulman2013} to solve this non-convex optimization problem. SCP solves a non-convex problem by repeatedly constructing a convex subproblem-an approximation to the problem around the current iterate $x$. A local convex approximation of the non-convex constraints is added along with a penalty co-efficient in the objective function. This subproblem can be efficiently solved using convex solvers and used to generate a step $\Delta x$ that makes progress on the original problem. The penalty co-efficient is then adjusted during the optimization to ensure that the constraint violation is driven to zero. For more details on the optimization procedure, we refer the interested readers to \cite{schulman2013}.

\begin{remark}
In our experience, solving the above optimization problem is very challenging even for relatively simpler NN structures  (like in \eqref{eqn:nnet}) because of highly nonlinear outputs of neural networks. Moreover, this complexity increases further with more complex network structures. This is another reason for choosing a simple NN structure in our analysis. 
\end{remark} 

\subsection{Linear Quadratic Regulator (LQR)} \label{secn:LQR}
Let us define the solution to \eqref{eq:trajopt} as ${\bf s}_*^{N_H} \coloneqq \{s_*(0),s_*(1),\hdots,s_*(N_H)\}$ and ${\bf u_*}^{N_H} \coloneqq \{u_*(0),u_*(1),\hdots,u_*(N_H-1)\}$. In practice, applying the control signal ${\bf u}^{N_H}_*$ could yield a trajectory that significantly differs from ${\bf s}^{N_H}_*$ due to (any remaining) mismatch between the model used in the optimization and actual dynamics of the quadrotor and unmodeled disturbances. This can be mitigated via feedback.
LQR is a well-known state feedback scheme for designing stabilizing controllers for trajectory tracking in linear systems. The technique can also be extended to nonlinear systems by linearizing the dynamics of the system about the desired trajectory. To stabilize the quadrotors on a (feasible) reference trajectory, we use a LQR feedback controller designed for the near hover model of quadrotor, along with a reference rotation before applying the feedback to correct for a non-zero yaw. 

A quadrotor is said to be in the \textit{hover} condition if the plane of the rotors is perpendicular to the vertical and it is at zero velocity relative to some inertial frame. The quadrotor position and orientation in space can be modified from the hover condition by varying the speeds of the motors from their hover speed. An approximate \textit{linear model} can be derived that accurately represents the quadrotor dynamics for small perturbations from the hover state \cite{bouffard2012board, beard2008quadrotor}. Since the quadrotor hover dynamics are derived around zero yaw, they no longer accurately represent the dynamics of the quadrotor for non-zero yaw states. In fact, there are highly nonlinear couplings between translational and rotational accelerations for non-zero yaw \cite{sanchez}. However, for any given non-zero yaw, one can consider a different inertial frame such that the yaw is zero with respect to the new inertial frame. In this inertial frame, the quadrotor is in near hover condition and hence one can use the LQR controller designed for the near hover model to control the system. We formalize this control scheme below.

Given a dynamically feasible reference trajectory ${\bf s}_*^{N_H}$, and nominal (open-loop) signal ${\bf u_*}^{N_H}$, define the error dynamics $\bar{s}(n)=s(n)-s_*(n)$ and compensation input $\bar{u}(n)=u(n)-u_*(n)$. Also, let the system dynamics be given by $s(n+1) = As(n)+Bu(n)$. The dynamics of the error signal can thus be obtained by:
\begin{align} \label{eq:err_dyn}
\bar{s}(n+1)=A\bar{s}(n)+B\bar{u}(n).
\end{align}
Subject to the dynamic constraints \eqref{eq:err_dyn}, the objective in LQR is to find a controller that minimizes the following quadratic cost starting from a given error state $\bar{s}_0$,
\begin{equation} \label{eqn:cost}
J_N(\bar{s}_0) =\sum_{n=0}^{N}\bar{s}(n)^TQ\bar{s}(n)+\bar{u}(n)^TR\bar{u}(n),
\end{equation}
\noindent where $Q$ and $R$ are positive definite matrices of appropriate size, and $\bar{s}(0)=\bar{s}_0$. The minimum cost to go $J^*_N$ can be solved recursively via dynamic programming, which also yields a time-invariant state feedback matrix $K$ that takes in the error state and outputs the appropriate compensation control. The closed loop control is thus given by
\begin{equation} \label{eqn:feedback_law}
u(n)=u_*(n) + K(s(n)-s_*(n)).
\end{equation} 
For our system, ${\bf s}_*^{N_H}, {\bf u}_*^{N_H}$ are obtained from SCP in Section \ref{secn:trajopt}. $A$ and $B$ matrices are obtained from the linear near-hover model proposed in \cite{bouffard2012board, beard2008quadrotor}. A feedback controller can thus be obtained by solving \eqref{eqn:cost} subject to \eqref{eq:err_dyn}, which is a convex-problem. However, when yaw is non-zero, the near hover model in \eqref{eq:err_dyn} is no longer valid and hence state feedback law given in \eqref{eqn:feedback_law} will no longer be able to stabilize the system. So at every time-step, we first rotate our inertial frame to another inertial frame in which yaw is zero. 

Let the error at step $n$ is given by $\bar{s}(n) := (\bar{p}(n), \bar{v}(n), \bar{\zeta}(n), \bar{\omega}(n))$, where $\bar{p}(n) = (\bar{x}(n), \bar{y}(n), \bar{z}(n))$ represents the error in position, and $(\bar{v}(n), \bar{\zeta}(n), \bar{\omega}(n))$ similarly represent errors in linear velocity, orientation and angular velocity respectively. Also, let the yaw at time $n$ be $\psi(n)$. The rotation of the inertial frame is thus equivalent to rotating the error in $(x,y)$ position and $(\dot{x}, \dot{y})$ by a rotation matrix as follows:
\begin{equation}
\begin{bmatrix} \bar{x}_R(n) \\ \bar{y}_R(n) \end{bmatrix} = T_R \begin{bmatrix} \bar{x}(n) \\ \bar{y}(n) \end{bmatrix}, ~~
\begin{bmatrix} \bar{\dot{x}}_R(n) \\ \bar{\dot{y}}_R(n) \end{bmatrix} = T_R \begin{bmatrix} \bar{\dot{x}}(n) \\ \bar{\dot{y}}(n) \end{bmatrix}, 
\end{equation}
where the rotation matrix is given by
\begin{equation}
T_R = \begin{bmatrix} cos(\psi(n)) & sin(\psi(n)) \\ -sin(\psi(n)) & cos(\psi(n))  \end{bmatrix}.
\end{equation}
The corresponding rotated error vector is $\bar{s}_R(n) := (\bar{p}_R(n), \bar{v}_R(n), \bar{\zeta}(n), \bar{\omega}(n))$, where $\bar{p}_R(n) = (\bar{x}_R(n), \bar{y}_R(n), \bar{z}(n))$ and $\bar{v}_R(n) = (\bar{\dot{x}}_R(n), \bar{\dot{y}}_R(n), \bar{\dot{z}}(n))$. Our overall state feedback control is thus given by:
\begin{equation} \label{eqn:feedback_law_final}
u(n)=u_*(n) + K\bar{s}_R(n).
\end{equation} 
\begin{remark}
Since the control law in \eqref{eqn:feedback_law_final} is derived using near-hover and only yaw rotation assumption, it is not the optimal feedback control when roll and pitch angles change significantly; however, this simple control scheme works well in practice and has been employed to fly an inverted pendulum in \cite{hehn2011flying} as well as by us in all our experiments. Nevertheless, more sophisticated control techniques can be developed based on the system dynamics \cite{sanchez}. 
\end{remark}
\begin{remark}
Note that the state feedback control law in \eqref{eqn:feedback_law_final} is good at error correction only when an accurate open-loop state ${\bf x}_*^N$ and control ${\bf u}_*^N$ are provided. Since the open-loop control depends heavily on the system model, the tracking with \eqref{eqn:feedback_law_final} can only be as good as the system dynamics model itself (for more details see Section \ref{secn:results}). In particular, for sinusoid-yaw reference trajectories, the NN should be able to learn the couplings between translational and rotational motion for a good tracking.
\end{remark}
\subsection{Crazyflie 2.0 and On-board PD Controller} \label{secn:PD_control}
To use the feedback control law in \eqref{eqn:feedback_law_final}, we need the full state $s(n)$. However, in practice, this information is obtained from different sensors which might run at different frequencies and hence the control update rate is limited by the frequency of the slowest sensor. This frequency, however, might not be enough to effectively control the system, and a low-level controller is thus required in practice to control the system between the two updates of the feedback loop. In this section, we first provide more details about our experiment testbed Crazyflie 2.0 and different sensors, and then design a low-level PD controller to control the system between the feedback updates. 

\begin{figure}
  \centering
  \includegraphics[width=0.45\textwidth]{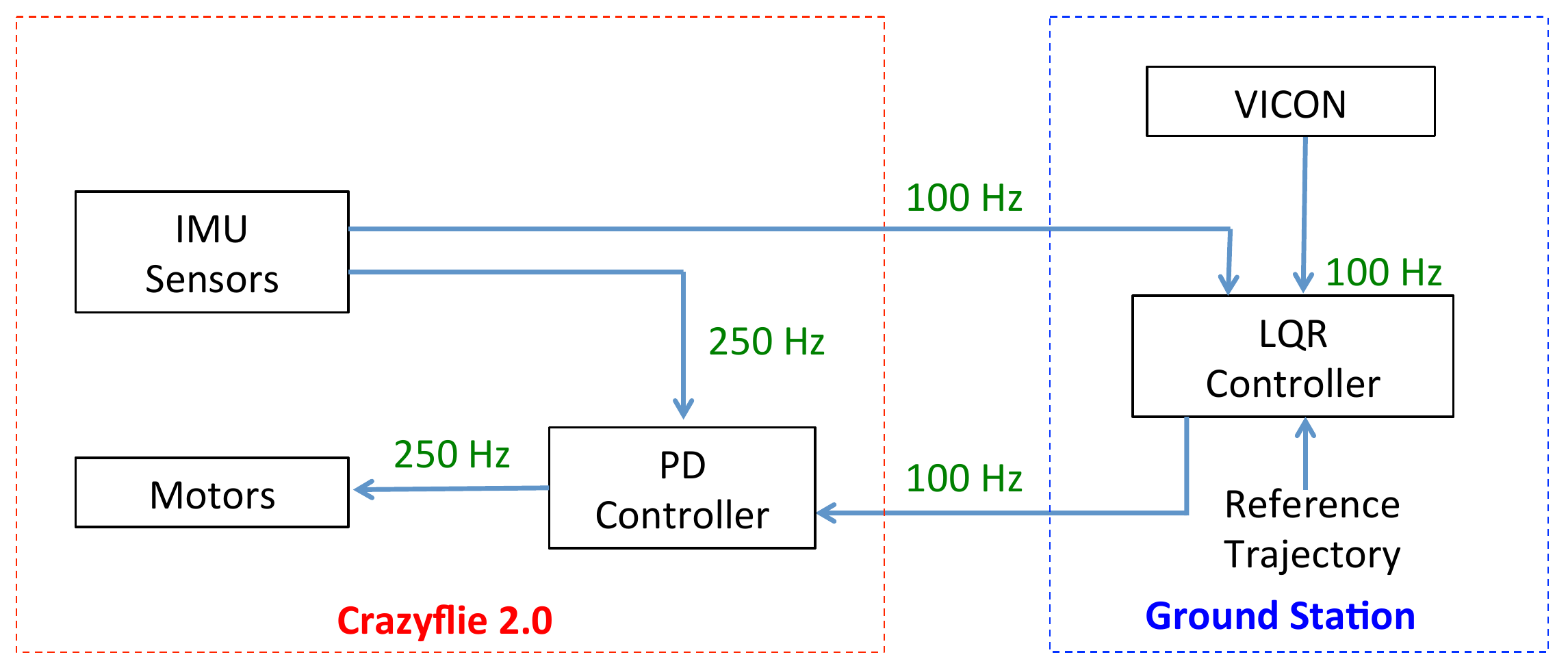}
  \caption{Control block diagram used to stabilize Crazyflie during experiments. At the ground station, LQR is running at 100Hz. On-board the Crazyflie, PD controller is running at 250Hz. Together they are able to stabilize the Crazyflie.}
    \vspace{-1em}
  \label{fig:control_scheme}
\end{figure}

The Crazyflie 2.0 is an open source nano quadrotor platform developed by Bitcraze. Its small size, low cost, and robustness make it an ideal platform for testing new control paradigms. Recently it has been used to exemplify aggressive flight in cluttered environments and for human robot interaction research \cite{landry2015, honig2015}. We use Crazyflie to collect training data as well as for the sinusoid-yaw experiments in this paper. 
We retrofit the quadrotor with reflective markers to allow for accurate position and velocity estimation via the VICON motion capture system at 100Hz. Furthermore, Crazyflie is equipped with an on-board inertial measurement unit (IMU) that provides orientation and angular velocity measurements at 250 Hz. VICON and IMU together thus provide the 12 dimensional state of the system and we can use the feedback control in \eqref{eqn:feedback_law_final} to stabilize the Crazyflie around the reference trajectory; however, our experiments indicate that this control scheme is not fast enough to keep the system stable. To overcome this problem, we implemented an on-board PD controller (proposed in \cite{landry2015}), which takes into account only the angular position and angular velocities that are available at a higher frequency of 250 Hz.
\begin{equation} \label{eq:pd}
	\begin{bmatrix} u_2 \\u_3\\ u_4 \end{bmatrix}=
	K_p\begin{bmatrix} \phi-\phi_{des} \\ \theta-\theta_{des} \\ \psi - \psi_{des} \end{bmatrix}+K_d\begin{bmatrix}  	\omega_x-\omega_{x, des} \\ \omega_y-\omega_{y, des} \\ \omega_z-\omega_{z, des}\end{bmatrix},
\end{equation}
\noindent where $K_p$ and $K_d$ are $3\times3$ matrices, $(\phi_{des}, \theta_{des}, \psi_{des})$ is the desired attitude, and $(\omega_{x, des}, \omega_{y, des}, \omega_{z, des})$ is the desired angular rate. Together LQR (100Hz) and PD controller (250Hz) are able to stabilize the Crazyflie around most of the trajectories. Note that we also did the same augmentation on our system in \eqref{sysDynamics} so that the new inputs to the system are now $\hat{u} := (u_1, \phi_{des}, \theta_{des}, \psi_{des}, \omega_{x, des}, \omega_{y, des}, \omega_{z, des})$, where mapping between inputs is given by \eqref{eq:pd}. The reference trajectory as well as the feedback law is thus computed for the augmented system. The full block diagram of our controller is shown in Figure \ref{fig:control_scheme}. 
\section{Experiments} \label{Exp}
In this section, we present the results of our experiments. We also discuss the data collection process to train the neural networks and the hyper-parameters used for training.

\subsection{Data Collection}
To collect data for training, we flew Crazyflie autonomously on a variety of trajectories, for example sinusoids in XY, XZ and YZ planes (but no yaw), and fixed position yaw-rotations, as well as manually on unstructured flights. For these flights, we recorded the state ($s$) and input ($\hat{u}$) data. Note that since we do not necessarily care about how closely we track a trajectory during the data collection process, 
the feedback controller discussed in Section \ref{secn:LQR} is sufficient to fly Crazyflie directly on the desired trajectories (that is, no feasible reference calculation is required); however, in general, the entire data can also be collected manually with experts flying the system. 

A picture of Crazyflie flying during one of our experiments is shown in Figure \ref{fig:crazyflie_pic}. One of our experimental videos can be found at: \textit{https://www.youtube.com/watch?v=QREeZvHg0lQ}. For communication between the ground station and Crazyflie, we used the Robot Operating System (ROS) framework \cite{quigley2009, Hoenig2015}. In total there are $2400$ seconds of flight time recorded, which correspond to $240,000$ $(s,\hat{u})$ data samples. Note that since we augment the system with a PD controller, we collect $\hat{u}$ during the flights, as opposed to $u$.   
\subsection{Neural Network Training} \label{secn:NN_training}
We next train two NNs (denoted as NN1 and NN2 here on) using the collected data to learn the linear and angular acceleration components $f_v$ and $f_{\omega}$ such that the MSE in \eqref{sysID_prob} is minimized. Before training the NNs, we follow a few data pre-processing steps: 
\begin{itemize}
\item Since we collect $\hat{u}$ (input to the PD-augmented system), we first derive $u$ (input to the quadrotor system in \eqref{sysDynamics}) from $\hat{u}$ using \eqref{eq:pd}, and use it as the input to the NNs along with the state information. This is to make sure that the learned system dynamics are independent of the control scheme. 
\item Instead of providing orientation angles as the input to the NNs, we provide sines and cosines of the angles. This is to make sure that NNs do not differentiate between $0$ and $2\pi$ radians.
\item We do not provide position as the input to any of the neural networks, as the translational and rotational accelerations should be position independent.
\item For each NN, we scale the observed outputs (for example, $x, y, z$ components of translational acceleration) such that each of them has zero mean and unity standard deviation. This is to make sure that the NNs give equal weightage to MSE in the three components.
\end{itemize}

The NN structure for each acceleration component is given by \eqref{eqn:nnet}. The objective (also called loss function) for each NN is to minimize the MSE between observed and predicted accelerations. The input to NN1 is $(v, \omega, sin(\zeta), cos(\zeta), u_1)$ and is $(v, \omega, sin(\zeta), cos(\zeta), u_2, u_3, u_4)$ for NN2. Note that we do not include $u_2, u_3, u_4$ in the input to NN1 because our experiments indicate that providing them result in over-fitting. Moreover, the physics of a quadrotor hints that the translational acceleration should not depend on these inputs \cite{beard2008quadrotor}. The same argument holds for $u_1$ and NN2. 

$60 \%$ of the total collected data was used for training, $25 \%$ was used for validation purposes and for tuning hyper parameters, and the rest was used for testing purposes. All the weights ($W, w$) and biases ($B, b$) were initially sampled from normal Gaussian distribution. For training the networks, we use the Neural Network Toolbox of MATLAB. We use the Resilient backpropagation learning algorithm. The learning rate, momentum constant, regularization factor and the number of hidden units were set at $0.01$, $0.95$, $0.1$ and $100$ respectively, but later tuned using the validation data. Overall, the learning algorithm makes about $100$ passes through the data, and optimize the weights and biases to minimize the loss function. Once the training is complete, the optimal weights and biases are obtained, which can be substituted in \eqref{eqn:nnet} to obtain the models for $f_v$ and $f_{\omega}$, and finally in \eqref{sysDynamics} to get the full dynamics model.

\begin{figure}
  \centering
  \includegraphics[width=0.45\textwidth]{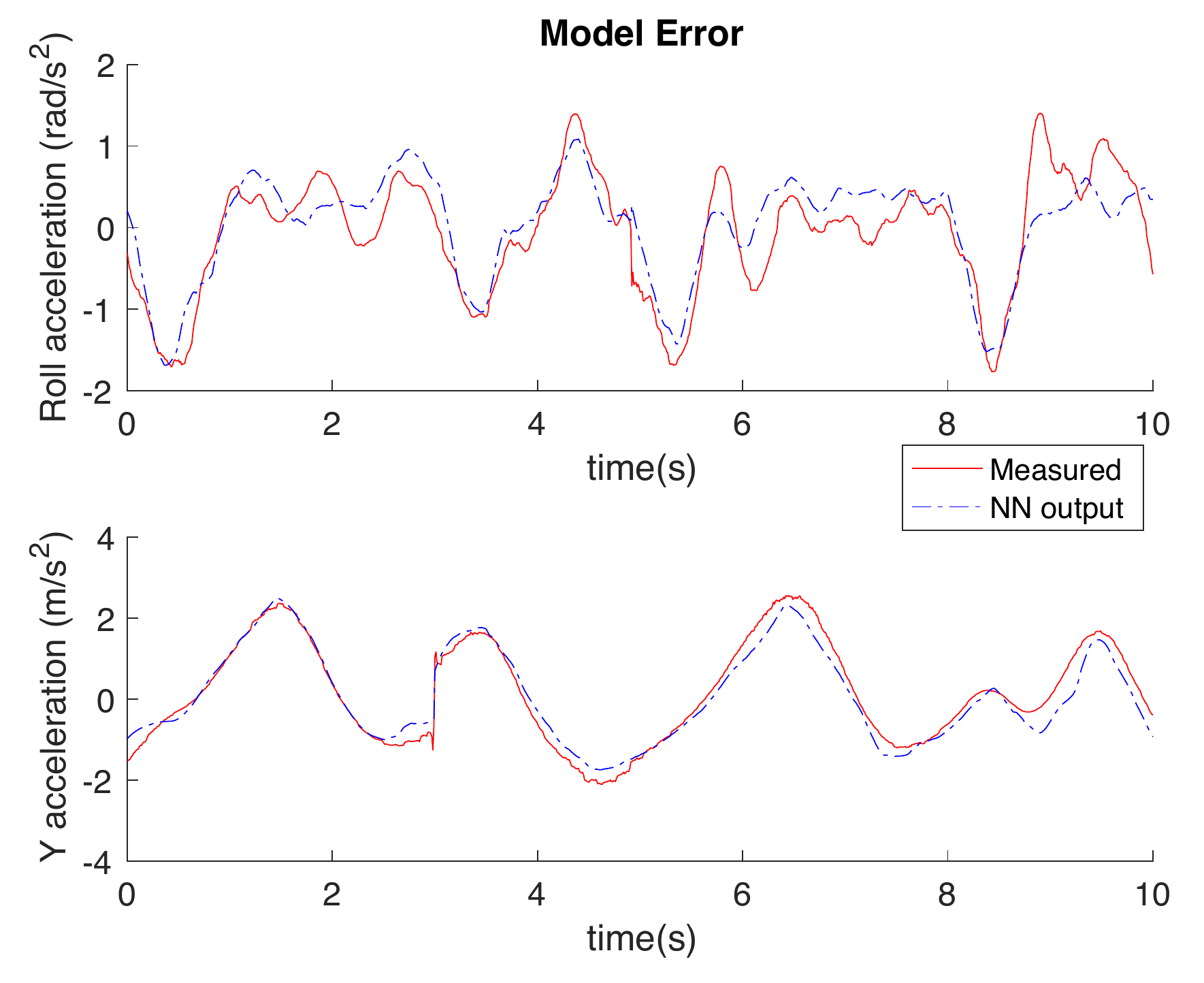}
  \caption{Observed and predicted values for the roll and y accelerations. The NNs are able to learn the acceleration models fairly accurately even with just the current state and input, indicating that the past states and inputs may not be required to learn the dynamics, and hence are avoided in this work to keep the control design simple.}
    \vspace{-2em}
  \label{fig:model_err}
\end{figure}

The (normalized) MSE numbers obtained for the training and the testing data after learning $f_v$ are $0.134$ and $0.135$ respectively, and that for $f_{\omega}$ are $0.341$ and $0.344$. Since the MSE numbers are very close for training and testing,  it indicates that the NNs do an accurate prediction on the unseen data as well, meaning that our NNs are not overfitting on the training data. In Figure \ref{fig:model_err}, we show the observed values and the predicted outputs of the trained NNs for roll and y accelerations. As evident from the figure, the NNs have been successfully able to learn the dynamics to a good accuracy. This indicates that a simple two-layer feed-forward NN structure used in this paper is sufficient to learn quadrotor dynamics to a good accuracy. Moreover, only the current state and input information is sufficient to learn the dynamics models, and hence past state and input information, which will potentially make the model as well as control design more complex, has not been used as an input to the NNs.    
\subsection{Sinusoid-yaw Trajectory Tracking Using NN Models} \label{secn:results}
Once NN1 and NN2 are trained, the full quadrotor model is available through \eqref{sysDynamics}. In this section, our goal is to use this model to track a sinusoid-yaw trajectory, where quadrotor is undergoing a sinusoidal motion in the XY plane while yawing at the same time. Since the desired trajectory consists of a simultaneous translational and rotational movement, the learned NN models should be able to capture the nonlinear couplings between these two movements to accurately track the trajectory. 

\begin{figure}
  \centering
  \includegraphics[width=0.45\textwidth]{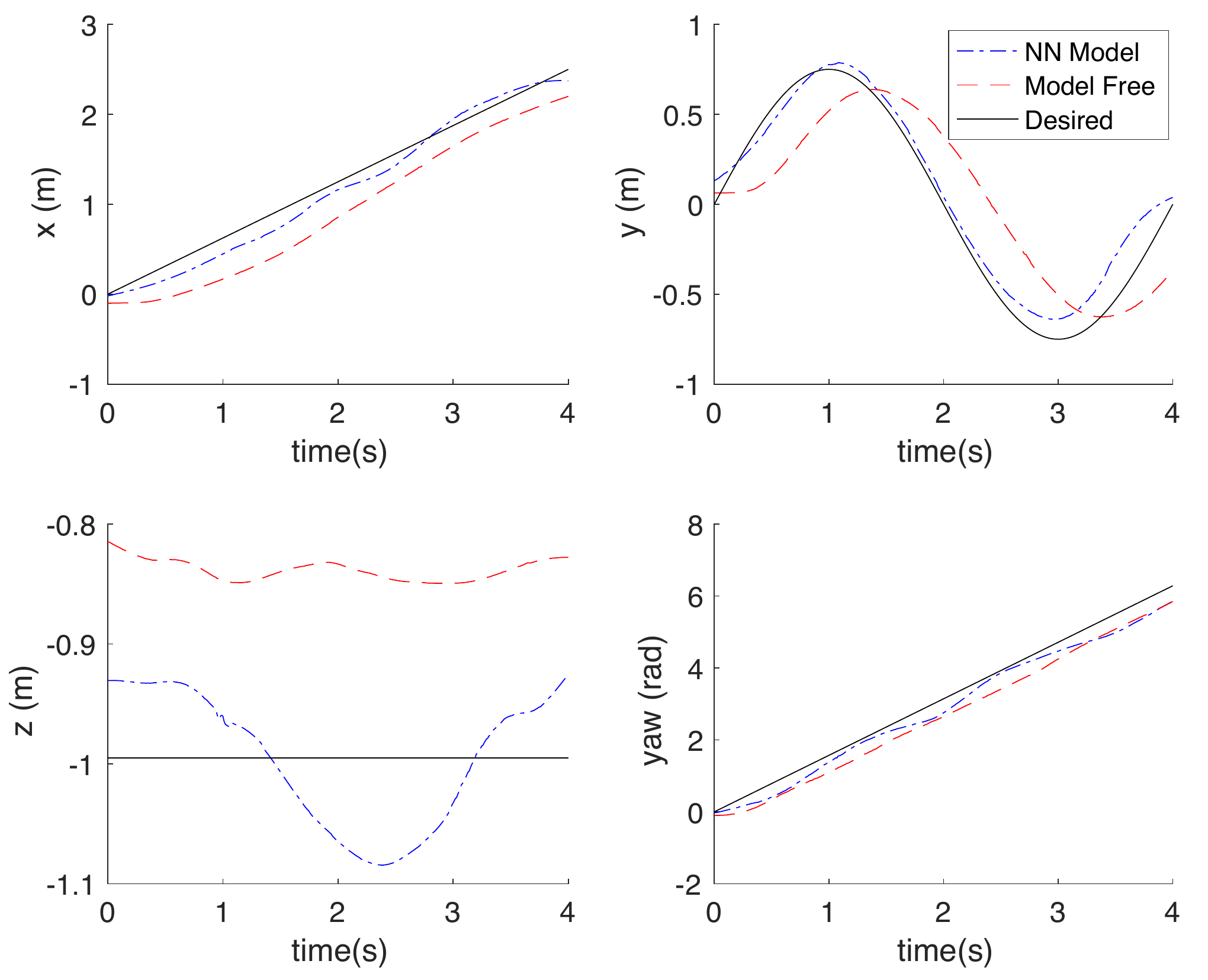}
  \caption{The reference, NN model and model-free trajectories obtained during the experiments. The NN model track the desired trajectory closely even though it involves both translational and rotational motion at the same time, which the NNs were not explicitly trained on, indicating the generalization capabilities of deep neural networks.}
  \vspace{-1em}
  \label{fig:tracking_figure}
\end{figure}

Using the full model, we first compute a dynamically feasible reference that is as close as possible to the desired sinusoid-yaw trajectory (shown in Figure \ref{fig:tracking_figure}), using the sequential convex programming method outlined in Section \ref{secn:trajopt}. The reference trajectory is then flown using the near-hover LQR scheme along with a yaw rotation as described in Section \ref{secn:LQR}. One of the tracking videos recorded during our experiments can be found at: \textit{https://www.youtube.com/watch?v=AeIfZbkjWPA}. We label the results corresponding to this experiment as `NN model' trajectory. 

The desired and NN model trajectories are shown in Figure \ref{fig:tracking_figure}. As evident from the figure, the NN model trajectory is able to track the desired trajectory closely. This illustrates that:
\begin{itemize}
\item the trained NNs are able to generalize the dynamics beyond the training data. In particular, the NN models capture the nonlinear couplings between translational and rotational accelerations, and can be used to track the trajectories they were not trained on.
\item even simple NN architectures, such as one used in this paper, have good generalization capabilities and can be used to control a quadrotor on complex trajectories. 
\end{itemize} 

From the results thus far it is not clear how much of the control performance is due to the open-loop signal derived from the NN model, since the LQR control may be correcting for model inconsistency. Therefore, we opted to fly Crazyflie using simply the LQR controller and desired trajectory as a reference and no open-loop control. 
We label the results corresponding to this experiment as the `model-free' trajectory. 

\begin{figure}
  \centering
  \includegraphics[width=0.45\textwidth]{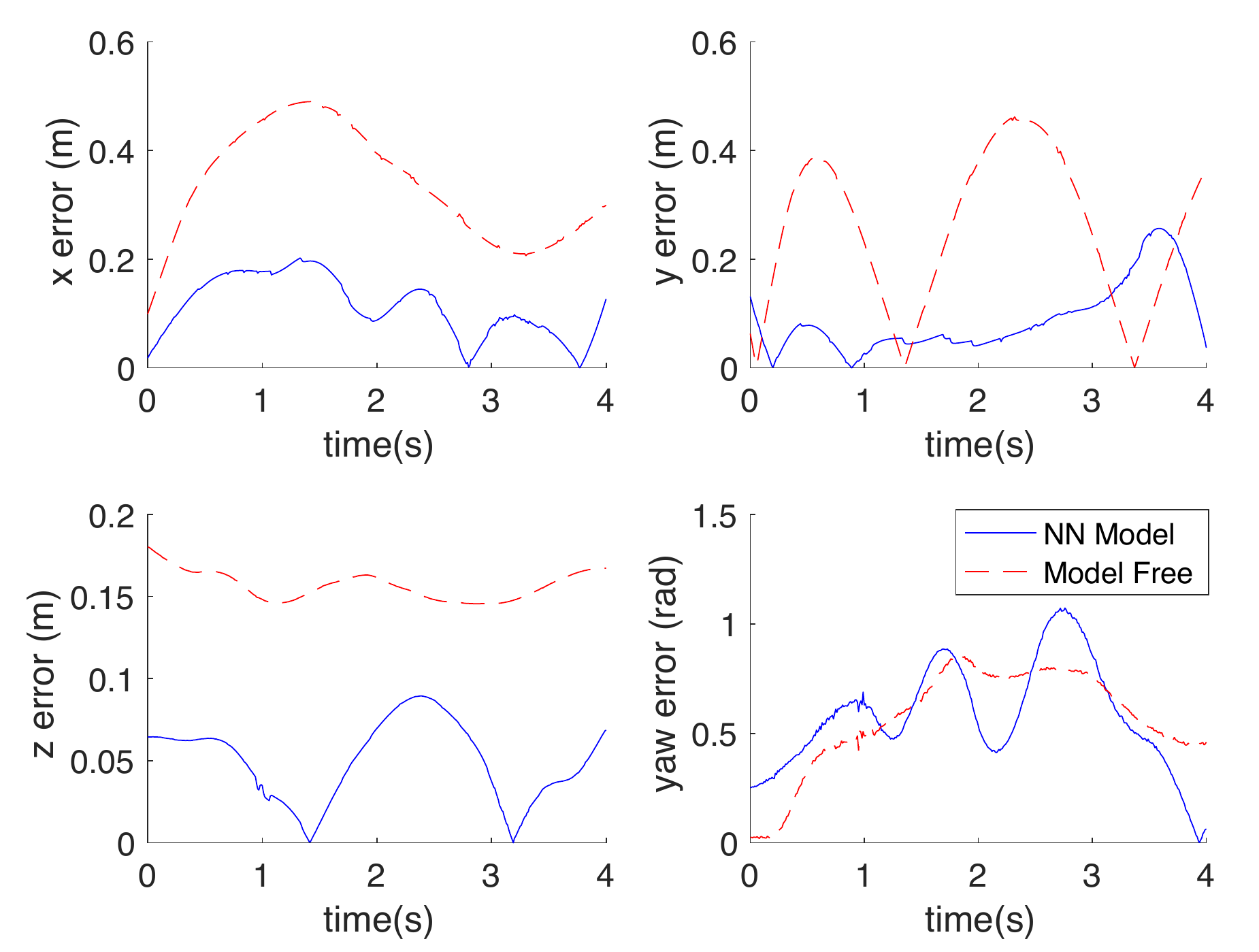}
  \caption{(Absolute) Tracking error for model-free and NN model trajectories. Model-free trajectory has a significantly higher tracking error compared to the NN model, especially in the translational motion, indicating that nonlinear coupling between translational and rotational motions should be taken into account while designing a controller, which in this work is captured by training a NN model that accurately represents the system dynamics.}
  \vspace{-1em}
  \label{fig:tracking_error}
\end{figure}

In Figure \ref{fig:tracking_error}, we show the (absolute) tracking error for the NN model and model-free trajectories. The NN model has a significantly lower tracking error compared to the model-free trajectory, indicating that the open-loop control derived from the NN model results in better tracking of the desired trajectory.
The reduced tracking error is thus the result of the availability of an accurate dynamics model. In general, to ensure a small tracking error, we need a good open-loop control, which in turn need a good dynamics model that accurately represents the system dynamics around the desired trajectory. 

Our experiments thus indicate that given the state-input data of a complex dynamical system (quadrotors in our case), deep neural networks are capable of learning the system dynamics to a good accuracy, and can represent the system behavior beyond the data they were trained on. Moreover, with a careful choice of the network architecture and its inputs, the NN model can be used effectively to control the system. Thus, deep neural networks seem to present a good alternative for the system identification of complex systems such as quadrotors, especially in scenarios in which it is hard to derive a physics-based model of the system.

\begin{remark}
Note that even though neural networks present one approach to identify the complex system dynamics, it is not the only approach. In our experience, one can use the general nonlinear model of quadrotor derived using Newtonian-Euler formalism \cite{beard2008quadrotor, hoffmann2007} along with the control scheme proposed in Section \ref{secn:CD} to get good tracking as well. Our goal in this paper, however, was to test the efficacy of neural network in learning a dynamic model, and not to compare different system models. In practice, one can use the physics-based model to collect data about the system and then use a neural network to learn incremental unmodeled dynamics (on top of the physics-based model) which might lead to an improved control performance. 
\end{remark}

\section{Conclusion} \label{done}
Traditional learning approaches proposed for controlling quadrotors have focused on improving the control performance for specific trajectories. In this work, we use deep neural networks to generalize the dynamics of the system beyond the trajectories used for training. Our experiments indicate that even simple NNs such as feed-forward networks can also have good generalization capabilities and can learn the dynamics of a quadrotor to good accuracy. More importantly, we demonstrate that the learned dynamics can be used effectively to control the system. Thus NNs are not only useful in being a good function approximator, but in this instance we can actually exploit the function that it produces for control purposes. For future work, it will be interesting to analyze whether combining a NN model with a physics-based model can lead to an improved control performance, thus utilizing both the known information about the system as well as the generalization capabilities of NNs.

\addtolength{\textheight}{1.5cm}   



%

%



\bibliographystyle{plain}
\bibliography{references}

\begin{thebibliography}{10}

\bibitem{abbeel2}
Pieter Abbeel, Adam Coates, and Andrew~Y Ng.
\newblock Autonomous helicopter aerobatics through apprenticeship learning.
\newblock {\em The International Journal of Robotics Research}, 2010.

\bibitem{beard2008quadrotor}
Randal Beard.
\newblock Quadrotor dynamics and control {Rev} 0.1.
\newblock 2008.

\bibitem{bouffard2012board}
Patrick Bouffard.
\newblock On-board model predictive control of a quadrotor helicopter: Design,
  implementation, and experiments.
\newblock Technical report, DTIC Document, 2012.

\bibitem{brookner1998tracking}
Eli Brookner.
\newblock Tracking and kalman filtering made easy, {John Wiley and Sons}.
\newblock {\em Inc. NY}, 1998.

\bibitem{hehn2011flying}
Markus Hehn and Raffaello D'Andrea.
\newblock A flying inverted pendulum.
\newblock In {\em Robotics and Automation (ICRA), 2011 IEEE International
  Conference on}, pages 763--770. IEEE, 2011.

\bibitem{hehn2013iterative}
Markus Hehn and Raffaello D’Andrea.
\newblock An iterative learning scheme for high performance, periodic
  quadrocopter trajectories.
\newblock In {\em European Control Conference (ECC). IEEE}, pages 1799--1804,
  2013.

\bibitem{Hoenig2015}
Wolfgang Hoenig, Christina Milanes, Lisa Scaria, Thai Phan, Mark Bolas, and
  Nora Ayanian.
\newblock Mixed reality for robotics.
\newblock In {\em IEEE/RSJ Intl Conf. Intelligent Robots and Systems}, pages
  5382 -- 5387, Hamburg, Germany, Sept 2015.

\bibitem{hoffmann2007}
Gabriel~M Hoffmann, Haomiao Huang, Steven~L Waslander, and Claire~J Tomlin.
\newblock Quadrotor helicopter flight dynamics and control: Theory and
  experiment.
\newblock In {\em Proc. of the AIAA Guidance, Navigation, and Control
  Conference}, volume~2, 2007.

\bibitem{honig2015}
Wolfgang Honig, Christina Milanes, Lisa Scaria, Thai Phan, Mark Bolas, and Nora
  Ayanian.
\newblock Mixed reality for robotics.
\newblock In {\em Intelligent Robots and Systems (IROS), 2015 IEEE/RSJ
  International Conference on}, pages 5382--5387. IEEE, 2015.

\bibitem{krizhevsky2012imagenet}
Alex Krizhevsky, Ilya Sutskever, and Geoffrey~E Hinton.
\newblock Imagenet classification with deep convolutional neural networks.
\newblock In {\em Advances in neural information processing systems}, pages
  1097--1105, 2012.

\bibitem{landry2015}
Benoit Landry.
\newblock Planning and control for quadrotor flight through cluttered
  environments.
\newblock Master's thesis, Massachusetts Institute of Technology, 2015.

\bibitem{lecun2015deep}
Yann LeCun, Yoshua Bengio, and Geoffrey Hinton.
\newblock Deep learning.
\newblock {\em Nature}, 521(7553):436--444, 2015.

\bibitem{nicol2011robust}
C~Nicol, CJB Macnab, and A~Ramirez-Serrano.
\newblock Robust adaptive control of a quadrotor helicopter.
\newblock {\em Mechatronics}, 21(6):927--938, 2011.

\bibitem{punjanideep}
Ali Punjani and Pieter Abbeel.
\newblock Deep learning helicopter dynamics models.
\newblock In {\em Robotics and Automation (ICRA), 2015 IEEE International
  Conference on}, pages 3223--3230. IEEE, 2015.

\bibitem{quigley2009}
Morgan Quigley, Ken Conley, Brian Gerkey, Josh Faust, Tully Foote, Jeremy
  Leibs, Rob Wheeler, and Andrew~Y Ng.
\newblock {ROS}: an open-source robot operating system.
\newblock In {\em ICRA workshop on open source software}, volume~3, page~5,
  2009.

\bibitem{raffo2010integral}
Guilherme~V Raffo, Manuel~G Ortega, and Francisco~R Rubio.
\newblock An integral predictive/nonlinear {$H_{\infty}$} control structure for
  a quadrotor helicopter.
\newblock {\em Automatica}, 46(1):29--39, 2010.

\bibitem{Rawlings2009}
J.~B. Rawlings and D.~Q. Mayne.
\newblock {\em Model predictive control: Theory and design}.
\newblock Nob Hill Pub., 2009.

\bibitem{sanchez}
Anand Sanchez-Orta, Vicente Parra-Vega, Carlos Izaguirre-Espinosa, and Octavio
  Garcia.
\newblock Position--yaw tracking of quadrotors.
\newblock {\em Journal of Dynamic Systems, Measurement, and Control},
  137(6):061011, 2015.

\bibitem{schulman2013}
John Schulman, Jonathan Ho, Alex~X Lee, Ibrahim Awwal, Henry Bradlow, and
  Pieter Abbeel.
\newblock Finding locally optimal, collision-free trajectories with sequential
  convex optimization.
\newblock In {\em Robotics: science and systems}, volume~9, pages 1--10.
  Citeseer, 2013.

\bibitem{voos2009nonlinear}
Ilolger Voos.
\newblock Nonlinear control of a quadrotor micro-{UAV} using
  feedback-linearization.
\newblock In {\em Mechatronics, 2009. ICM 2009. IEEE International Conference
  on}, pages 1--6. IEEE, 2009.

\bibitem{wan2000unscented}
Eric~A Wan and Ronell Van Der~Merwe.
\newblock The unscented kalman filter for nonlinear estimation.
\newblock In {\em Adaptive Systems for Signal Processing, Communications, and
  Control Symposium 2000. AS-SPCC. The IEEE 2000}, pages 153--158. Ieee, 2000.

\bibitem{xu2006sliding}
Rong Xu and {\"U}mit {\"O}zg{\"u}ner.
\newblock Sliding mode control of a quadrotor helicopter.
\newblock In {\em Decision and Control, 2006 45th IEEE Conference on}, pages
  4957--4962. IEEE, 2006.

\bibitem{zhaowei2015iterative}
Ma~Zhaowei, Hu~Tianjiang, Shen Lincheng, Kong Weiwei, Zhao Boxin, and Yao
  Kaidi.
\newblock An iterative learning controller for quadrotor {UAV} path following
  at a constant altitude.
\newblock In {\em Control Conference (CCC), 2015 34th Chinese}, pages
  4406--4411. IEEE, 2015.

\end{thebibliography}

\end{document}